# Higher-order Landau-Devonshire theory for BaTiO$_3$

Ilhwan Kim[1], Kumok Jang[1], Ilhun Kim[1, 2], Lin Li[2]

[1] Department of Physics, Kim Hyong Jik Normal University, Pyongyang, Democratic People's Republic of Korea

[2] College of Science, Northeastern University, Shenyang 110819, People's Republic of China

Corresponding author: ririn@sohu.com (L. Li)

The higher-order Landau-Devonshire theory for BaTiO$_3$ is proposed. The structural stabilities of some Landau potentials proposed for the phenomenology of BaTiO$_3$ until now are discussed in the framework of the singularity theory. We confirm that the structurally stable Landau potential has to contain at least all the invariants up to eighth power when the three parameters vary in the experiment. We propose the tenth order Landau potential for adequate description of the various experimental data for BaTiO$_3$ crystal. We show that the results of the phenomenology based on the tenth potential are in good agreement with experimental data on the spontaneous polarization, dielectric constants, dielectric susceptibility and piezoelectric coefficients versus the temperature and the electric field.

## I. INTRODUCTION

The classic Landau-Devonshire theory gave a natural explanation for the appearance of tetragonal, orthorhombic, and rhombohedral phases in materials, such as BaTiO$_3$, KNbO$_3$ and PbTiO$_3$, on the basis of the fourth-order and the sixth-order free-energy expansion with only a single temperature dependent second-order coefficient, and the other temperature independent higher-order coefficients.

Recent x-ray and neutron diffraction studies and first-principles calculations on the perovskite solid solutions (1-$x$)PbZrO$_3$-$x$PbTiO$_3$(PZT), (1−$x$)Pb(Mg$_{1/3}$Nb$_{2/3}$)O$_3$-$x$PbTiO$_3$(PMN-$x$PT), (1−$x$)Pb(Mg$_{1/3}$Nb$_{2/3}$)O$_3$-$x$PbTiO$_3$(PZN-$x$PT) have revealed monoclinic, triclinic and orthorhombic phases *Cm*, *B*2/*m*, *P*1, *Amm*2 in the vicinity of the MPB.

It was confirmed that the sixth-order Devonshire expansion does not allow for the occurrence of a monoclinic phase, and later higher-order Landau-Devonshire theory based on the potential containing the eighth, tenth and twelfth power terms of order parameter has been developing for description on the presence of newly observed low-symmetry phases.[1-4]

On the other hand, the dielectric constant, spontaneous polarization and piezoelectric coefficients of BaTiO$_3$, KNbO$_3$, PbTiO$_3$ single crystals, bulk ceramics, thin films and various materials in the nanometer size range can be precisely reproduced within the framework of the higher-order Landau-Devonshire theory.[5-13]

All these facts indicate that the eighth or higher power terms in the Landau potential expansion which were ignored in the traditional phenomenological framework play the essential role for adequate description of the thermodynamic behavior of the systems as well as the symmetry of the system.

When one constructs the phenomenological model of the system with the multi-component order parameters, it is very important to choose the simplified Landau potential model whose singularity is preserved.[14-16]

The singularity theory[17-19] (it is known as the catastrophe theory in physics) became well known as the powerful mathematical tool for construction of the structural stable phenomenological model. According to this theory, only the structural stable thermodynamic potential can describe the qualitative characteristics of the system.

Although the importance of application of the methods of singularity theory to the phenomenological investigation is early known, a few attentions has been paid to the structural stability of the phenomenological model of the phase transitions.

The structural stable twelfth order expansion of the Landau potential was only applied to description of the lower symmetry phases such as monoclinic and triclinic phases in the perovskite ferroelectric solid solutions, but no attention has been given to the structural stability of the high order potential for explaining highly nonlinearity of the physical properties in the various crystals.

An example of such a phenomenological theory of the ferroelectric phase transitions is one of $BaTiO_3$ crystal.

Since the discovery of ferroelectric $BaTiO_3$ more than 70 years ago, it has been attracted great interest due to its potential applications in the microelectronic device and become fine theoretical model to clarify the physical nature of the ferroelectric properties.

Since the phenomenological Landau-Ginzburg-Devonshire theory of the phase transition in $BaTiO_3$ constructed in the 1940s, it has been applying successfully for explaining interesting physical properties observed in ferroics and multiferroics such as ferroelectrics, ferromagnetics, ferroelastics and ferroelectric-ferromagnetics, and superconductors.

To reproduce the structure and ferroelectric properties including the phase transition temperature of $BaTiO_3$ single crystal and ceramic, A. J. Bell and L. E. Cross[5] proposed the Landau-Devonshire sixth order potential as polynomial of the ferroelectric order parameter-polarization $p$;

$$\Phi_1(\eta) = \alpha_1(p_1^2 + p_2^2 + p_3^2) + \alpha_{11}(p_1^4 + p_2^4 + p_3^4) + \\ + \alpha_{12}(p_1^2 p_2^2 + p_2^2 p_3^2 + p_1^2 p_3^2) + \alpha_{111}(p_1^6 + p_2^6 + p_3^6) + \\ + \alpha_{112}(p_1^2(p_3^4 + p_2^4) + p_2^2(p_1^4 + p_3^4) + p_3^2(p_1^4 + p_2^4)) + \\ + \alpha_{123} p_1^2 p_2^2 p_3^2 \quad (1)$$

where coefficients $\alpha_1$, $\alpha_{11}$ and $\alpha_{111}$ depend on the temperature, while the others are constants.

Potential model (1) has been used to predict the effect of strain on the phase transitions and ferroelectric properties of $BaTiO_3$ thin film(in this case $\alpha_1$, $\alpha_{11}$ and $\alpha_{123}$ depend on the temperature.[20]) as well as bulk. However, it is only applicable to $BaTiO_3$ thin film under relatively small compressive strains ($\leq 0.4\%$) and cannot explain the ferroelectric properties under large compressive constraints.[21, 22]

It is recently discovered that $BaTiO_3$ films can be compressively strained as much as 1.6%, and the cubic to tetragonal ferroelectric transition temperature of $BaTiO_3$ thin films can be increased to over 600 ℃[23], a huge shift compared with bulk $BaTiO_3$ single crystal(~125 ℃). A thermodynamic potential was required to predict the phase transition, domain structure and ferroelectric properties of $BaTiO_3$ thin films under such large compressive constraints.

Y. L. Li et al.[7] suggested the Landau-type potential up to eighth order in terms of the polarization $p$;

$$\Phi_2(\eta) = \alpha_1(p_1^2 + p_2^2 + p_3^2) + \alpha_{11}(p_1^4 + p_2^4 + p_3^4) + \\ + \alpha_{12}(p_1^2 p_2^2 + p_2^2 p_3^2 + p_1^2 p_3^2) + \alpha_{111}(p_1^6 + p_2^6 + p_3^6) + \\ + \alpha_{112}[p_1^2(p_3^4 + p_2^4) + p_2^2(p_1^4 + p_3^4) + p_3^2(p_1^4 + p_2^4)] + \\ + \alpha_{123} p_1^2 p_2^2 p_3^2 + \alpha_{1111}(p_1^8 + p_2^8 + p_3^8) + \\ + \alpha_{1112}[p_1^6(p_3^2 + p_2^2) + p_2^6(p_1^2 + p_3^2) + p_3^6(p_1^2 + p_2^2)] + \\ + \alpha_{1122}(p_1^4 p_2^4 + p_2^4 p_3^4 + p_1^4 p_3^4) + \\ + \alpha_{1123}(p_1^4 p_2^2 p_3^2 + p_2^4 p_1^2 p_3^2 + p_3^4 p_1^2 p_2^2) \quad (2)$$

where all coefficients are assumed to be temperature independent except $\alpha_1$.

The polarization vector, dielectric constants, ferroelectric transition temperature and electric field dependence of piezoelectric coefficients of bulk $BaTiO_3$ single crystal which are calculated by potential model (2) are in good agreement with experimental data.

After Li et al.'s work, the higher-order Landau-Devonshire theory for various ferroelectric thin film and bulk materials was constructed on the basis of the eighth order potential model (2). [8-12, 24-26]

After comparing Bell-Cross's model[5] and N. Pertsev's one[20] with three variable parameters and Y. L. Li's model[7] with one variable parameter, and investigating the dielectric properties of BaTiO$_3$ single crystal, Y. L. Wang et al.[8] stressed that not only the eighth-power invariants but also four temperature dependent anharmonic coefficients $\alpha_1$, $\alpha_{11}$, $\alpha_{111}$ and $\alpha_{12}$ are also necessary for more correct quantitative description of the experimental data related to ferroelectric properties of BaTiO$_3$.

Having compared the experimental data on the spontaneous polarization, dielectric constant and piezoelectric coefficients under superstrong electric field value up to 55MV/m with the results of the phenomenological Landau–Devonshire approach, I. B. Leontyev et al.[27] showed that either the potential model by Bell et al.[6] or one by Li et al.[7] explains better his experimental data than Y. L. Wang's one.

Moreover, J. J. Wang et al.[12] assumed that six coefficients in the expansion (2) depend on the external parameters in order to explain experimental data on the temperature-pressure phase diagram of BaTiO$_3$ crystal. Among these coefficients, $\alpha_1$ depend on the temperature, while $\alpha_{11}$, $\alpha_{111}$, $\alpha_{12}$, $\alpha_{112}$ and $\alpha_{123}$- pressure.

X. Lu et al.[13] estimate temperature dependence of the coefficients in the several sixth and eighth order potential models of BaTiO$_3$ on the basis of data with respect to phase transitions induced by electric field (double hysteresis loop) and showed that the expansion coefficients proposed by Bell and Cross[5] are adequate for the bulk crystal and the ones by Petsev et al.[20] - thin film.

On the other hand, according to first principles calculation[28], coefficients of the quadratic and two fourth power terms in the Landau–Devonshire potential for BaTiO$_3$ have to be temperature dependent.

According to singularity theory[16-19], the singularities of the Landau potentials for description of the phase transition in the crystal are determined by the symmetry of the order parameters and the dependence of the phenomenological coefficients on the external parameters such as temperature, pressure and concentration.

However, until now the potential models widely used in the phenomenological study of BaTiO$_3$ crystal were constructed by the perturbation theory of the traditional Landau theory, no one paid attention to investigate the structural stability of the potential on the base of the singularity theory.

In this paper, the structural stabilities of the Landau phenomenological models proposed for BaTiO$_3$ crystal are discussed and new structural stable phenomenological models will be suggested. Also the temperature and the electric field dependence of the ferroelectric properties of BaTiO$_3$ will be calculated using the structural stable phenomenological model.

## II. STRUCTURAL STABILITY OF PHENTMENOLOGICAL MODEL INVAIANT UNDER CUBIC POINT GROUP $m\bar{3}m$

### A. Structural stability of phenomenological models

In the ABO$_3$-type ferroelectrics such as BaTiO$_3$, PbTiO$_3$, PbZr$_{1-x}$Ti$_x$O$_3$(PZT) with the space group $Pm\bar{3}m$ of the prototype phase, the successive ferroelectric phase transitions $Pm\bar{3}m - P4mm - Amm2 - R3m$ are observed as a result of soft mode instability at the center($\Gamma$-point) of primitive cubic Brillouin zone.

The sequence of the ferroelectric transitions can be explained by Landau-Devonshire potential in a power expansion of following three basis invariants of the image group I= $m\bar{3}m$.[15, 29, 30]

$$J_1 = p_1^2 + p_2^2 + p_3^2,$$
$$J_2 = p_1^2 p_2^2 + p_2^2 p_3^2 + p_3^2 p_1^2, \quad (3)$$
$$J_3 = p_1^2 p_2^2 p_3^2.$$

The invariant polynomials $J_1$, $J_2$ and $J_3$ then become the basis functions in the orbit space

$W = \varepsilon_3 / I\varepsilon_3$ generated by action of the cubic point group $m\bar{3}m$ on the three dimensional order parameter space $\varepsilon_3$.

The Landau-Devonshire potential invariant under the cubic point group $m\bar{3}m$ is expressed in the orbit space W.[14, 16]

$$\begin{aligned}\Phi = \Phi_0 &+ a_1 J_1 + a_2 J_2 + a_{11} J_1^2 + a_3 J_3 + a_{12} J_1 J_2 + a_{111} J_1^3 + \\
&+ a_{1111} J_1^4 + a_{112} J_1^2 J_2 + a_{13} J_1 J_3 + a_{22} J_2^2 + a_{11111} J_1^5 + \\
&+ a_{1112} J_1^3 J_2 + a_{113} J_1^2 J_3 + a_{122} J_1 J_2^2 + a_{23} J_2 J_3 + \\
&+ a_{111111} J_1^6 + a_{222} J_2^3 + a_{33} J_3^2 + a_{11112} J_1^4 J_2 + a_{1113} J_1^3 J_3 + \\
&+ a_{1122} J_1^2 J_2^2 + a_{123} J_1 J_2 J_3 + ...\end{aligned} \quad (4)$$

According to the singularity theory, action of the equivariant vector field

$$U_k = \sum_{i,m} (\nabla_i J_k, \nabla_i J_m) \frac{\partial}{\partial J_m} \quad (5)$$

on the potential (4) gives the gradient ideal $I_{\nabla \Phi}$, and the elements of $I_{\nabla \Phi}$ don't affect the structural stability of the potential (4). Consequently, the terms written as

$$I_{\nabla \Phi} = U_k \Phi = \sum_{i,m} (\nabla_i J_k, \nabla_i J_m) \frac{\partial \Phi}{\partial J_m} \quad (6)$$

are the perturbation terms able to be dropped near the critical point.

Now, let's consider the structural stability of the Landau-Devonshire potentials widely applied to the phenomenology of BaTiO$_3$ crystal.

Considering the potential (4), Bell et al.'s sixth order expansion (1) is expressed in terms of basis invariant polynomials (3):

$$\Phi(J) = a_1 J_1 + a_{11} J_1^2 + a_2 J_2 + a_{111} J_1^3 + a_{12} J_1 J_2 + a_3 J_3, \quad (7)$$

where $a_1 = \alpha_1$, $a_{11} = \alpha_{11}$, $a_2 = \alpha_{12}$, $a_{111} = \alpha_{111}$, $a_{12} = \alpha_{112}$, $a_3 = \alpha_{123}$ and expansion coefficients $a_1$, $a_{11}$ and $a_{111}$ are assumed to be temperature dependent.

To show, if the phenomenological model (7) is structurally stable under the change in the coefficients $a_1$, $a_{11}$ and $a_{111}$ with temperature, we have to find the universal unfolding[19] under control parameters $a_1$, $a_{11}$ and $a_{111}$.

According to the formula (5), the equivariant vector field with respect to basis invariants (3) is given by

$$\begin{aligned}U_1 &= J_1 \frac{\partial}{\partial J_1} + J_2 \frac{\partial}{\partial J_2} + J_3 \frac{\partial}{\partial J_3}, \\
U_2 &= J_2 \frac{\partial}{\partial J_1} + (J_1 J_2 + J_3) \frac{\partial}{\partial J_2} + J_1 J_3 \frac{\partial}{\partial J_3}, \\
U_3 &= J_3 \frac{\partial}{\partial J_1} + J_1 J_3 \frac{\partial}{\partial J_2} + J_2 J_3 \frac{\partial}{\partial J_3}.\end{aligned} \quad (8)$$

Now, using the formula (8), we can obtain gradient ideal $I_{\nabla \Phi}$ and the factor of the ring of polynomials $\Phi(J)$ over the gradient ideal $I_{\nabla \Phi}$:

$$Q = \Phi(J) / I_{\nabla \Phi}$$

The singularity of the potential $\Phi(J)$ is determined by the $Q$-local algebra[16, 17]:

$$Q = \{J_1, J_1^2, J_1^3, J_1^4, J_1^5, J_2\} \quad (9)$$

The universal unfolding of the singularity of $\Phi(J)$ with the control parameters $a_1$, $a_{11}$ and $a_{111}$, that is structural stable potential, is the family of polynomials in which enter all the monomials of the $Q$-local algebra of the singularity.

From the formula (9), we can see that the structural stable potential has to be tenth order expansion in terms of the order parameter, but not sixth order one. Hence, it results in the fact that Bell-Cross's sixth order expansion (7) with temperature dependent coefficients $a_1$, $a_{11}$ and $a_{111}$ is structurally unstable.

On the other hand, sixth order potential model by Pertsev et al. corresponds to the potential (7) with temperature dependent $a_1$, $a_{11}$ and $a_3$.

In the case of control parameters $a_1$, $a_{11}$ and $a_3$, we have local algebra of the singularity of the potential $\Phi$:

$$Q = \{J_1, J_1^2, J_1^3, J_1^4, J_2, J_3\} \quad (10)$$

We can see that the structural stable potential model has to be expanded up to eighth order terms, but not sixth order terms. Therefore, sixth order potential model by Pertsev et al. with the temperature dependent coefficients $a_1$, $a_{11}$ and $a_3$ is structurally unstable.

Let's consider the eighth order expansion (2) by Li et al.. Potential model (2) is written in terms of the basis invariants (3):

$$\Phi(J) = a_1 J_1 + a_2 J_2 + a_{11} J_1^2 + a_3 J_3 + a_{12} J_1 J_2 + a_{111} J_1^3 + \\ + a_{1111} J_1^4 + a_{112} J_1^2 J_2 + a_{13} J_1 J_3 + a_{22} J_2^2 \quad (11)$$

assuming $a_1$ is temperature dependent.

In the case of single control parameter $a_1$, the model (11) is, of course, structurally stable because the universal unfolding is determined by $Q$-local algebra:

$$Q = \{J_1, J_1^2, J_2, J_3\}. \quad (12)$$

Therefore, the potential model (11) is structurally stable. But the qualitative characteristic of the system with one control parameter $a_1 = a_0(T - T_0)$ can be well explained via potential model

$$\Phi(J) = a_1 J_1 + a_{11} J_1^2 + a_2 J_2 + a_3 J_3, \quad (13)$$

instead of potential (11).

The Y. L. Wang et al.'s eighth order potential is the same as Li et al.'s eighth order one, with $a_1$, $a_{11}$, $a_{111}$ and $a_2$ as the variable parameters.

In the case of the control parameters $a_1$, $a_{11}$, $a_{111}$ and $a_2$, the local algebras are able to be obtained in three form as follows:

$$Q_1 = \{J_1, J_1^2, J_1^3, J_1^4, J_1^5, J_2, J_3, J_1 J_2, J_1 J_3\} \\ Q_2 = \{J_1, J_1^2, J_1^3, J_1^4, J_1^5, J_2, J_3, J_1 J_2, J_2^2\} \quad (14) \\ Q_3 = \{J_1, J_1^2, J_1^3, J_1^4, J_1^5, J_2, J_3, J_1 J_2, J_1^2 J_2\}$$

As shown in the formula (14), when $a_1$, $a_{11}$, $a_{111}$ and $a_2$ are chosen as the control parameters, structurally stable potential model has to contain tenth power invariants. Thus, the potential model by Y. L. Wang et al. is structurally unstable.

If the coefficients $a_1$, $a_2$, $a_{11}$, $a_{111}$, $a_{12}$ and $a_3$ in the formula (4) are variable, the structurally stable potential model is twelfth order expansion (Table 1), so eighth order potential model by J. J. Wang et al. is also structurally unstable.

It follows from the singularity theory that the potential models by A. J. Bell et al., N. A. Pertsev et al., Y. L. Wang et al. and J. J. Wang et al. are structurally unstable. In fact, the results found within the framework of the structurally unstable potential model could be nonphysical.[16-19]

## B. New Landau potential models

The study on the singularity of the phenomenological models invariant under cubic point group $m\bar{3}m$ widely used in the investigation of the physical properties of the solids results in the problem of finding all the universal unfoldings with co-rank 3 and various co-dimensions. Taking into account the results of previous theoretical[2-13] and experimental[31, 32] works, we choose the co-dimensions 1~4.

Using the singularity theory, we have obtained the universal unfoldings with respect to various control parameters. (Table I)

TABLE I. The universal unfoldings invariant under cubic point group $m\bar{3}m$

| Control parameters | universal unfoldings $\Phi(J)$ | The highest power |
|---|---|---|
| $a_1$ | $\Phi_1 = a_1 J_1 + a_{11} J_1^2 + a_2 J_2 + a_3 J_3$ | 6 |
| $a_1$, $a_{11}$ | $\Phi_2 = a_1 J_1 + a_{11} J_1^2 + a_{111} J_1^3 + a_{1111} J_1^4 + a_2 J_2$ | 8 |
| $a_1$, $a_2$ | $\Phi_3 = a_1 J_1 + a_{11} J_1^2 + a_2 J_2 + a_{22} J_2^2 + a_3 J_3 + a_{23} J_2 J_3$ | 10 |
| $a_1$, $a_3$ | $\Phi_4 = a_1 J_1 + a_{11} J_1^2 + a_2 J_2 + a_3 J_3$ | 6 |
| $a_1$, $a_2$, $a_{11}$ | $\Phi_5 = a_1 J_1 + a_{11} J_1^2 + a_{111} J_1^3 + a_2 J_2 + a_3 J_3 + a_{12} J_1 J_2 + a_{22} J_2^2 + a_{112} J_1^2 J_2$ | 8 |
|  | $\Phi_6 = a_1 J_1 + a_{11} J_1^2 + a_{111} J_1^3 + a_2 J_2 + a_3 J_3 + a_{12} J_1 J_2 + a_{22} J_2^2 + a_{13} J_1 J_3$ | 8 |
|  | $\Phi_7 = a_1 J_1 + a_{11} J_1^2 + a_{111} J_1^3 + a_2 J_2 + a_3 J_3 + a_{12} J_1 J_2 + a_{112} J_1^2 J_2 + a_{13} J_1 J_3$ | 8 |
|  | $\Phi_8 = a_1 J_1 + a_{11} J_1^2 + a_{111} J_1^3 + a_2 J_2 + a_3 J_3 + a_{12} J_1 J_2 + a_{1111} J_4^2 + a_{13} J_1 J_3$ | 8 |
|  | $\Phi_9 = a_1 J_1 + a_{11} J_1^2 + a_{111} J_1^3 + a_2 J_2 + a_3 J_3 + a_{12} J_1 J_2 + a_{1111} J_4^2 + a_{112} J_1^2 J_2$ | 8 |
| $a_1$, $a_2$, $a_3$ | $\Phi_{10} = a_1 J_1 + a_{11} J_1^2 + a_2 J_2 + a_{22} J_2^2 + a_3 J_3 + a_{33} J_3^2 + a_{23} J_2 J_3$ | 12 |
| $a_1$, $a_{11}$, $a_{111}$ | $\Phi_{11} = a_1 J_1 + a_{11} J_1^2 + a_{111} J_1^3 + a_{1111} J_1^4 + a_{11111} J_1^5 + a_2 J_2$ | 10 |
| $a_1$, $a_3$, $a_{11}$ | $\Phi_{12} = a_1 J_1 + a_{11} J_1^2 + a_{111} J_1^3 + a_{1111} J_1^4 + a_2 J_2 + a_3 J_3$ | 8 |
| $a_1$, $a_2$, $a_3$, $a_{11}$ | $\Phi_{13} = a_1 J_1 + a_{11} J_1^2 + a_{111} J_1^3 + a_2 J_2 + a_3 J_3 + a_{12} J_1 J_2 + a_{13} J_1 J_3 + a_{112} J_1^2 J_2 + a_{33} J_3^2$ | 12 |
|  | $\Phi_{14} = a_1 J_1 + a_{11} J_1^2 + a_{111} J_1^3 + a_{1111} J_1^4 + a_2 J_2 + a_3 J_3 + a_{12} J_1 J_2 + a_{22} J_2^2 + a_{33} J_3^2$ | 12 |
|  | $\Phi_{15} = a_1 J_1 + a_{11} J_1^2 + a_{111} J_1^3 + a_2 J_2 + a_3 J_3 + a_{12} J_1 J_2 + a_{22} J_2^2 + a_{112} J_1^2 J_2 + a_{33} J_3^2$ | 12 |
| $a_1$, $a_2$, $a_{11}$, $a_{111}$ | $\Phi_{16} = a_1 J_1 + a_{11} J_1^2 + a_{111} J_1^3 + a_{1111} J_1^4 + a_{11111} J_1^5 + a_2 J_2 + a_3 J_3 + a_{12} J_1 J_2 + a_{13} J_1 J_3$ | 10 |
|  | $\Phi_{17} = a_1 J_1 + a_{11} J_1^2 + a_{111} J_1^3 + a_{1111} J_1^4 + a_{11111} J_1^5 + a_2 J_2 + a_3 J_3 + a_{12} J_1 J_2 + a_{22} J_2^2$ | 10 |
|  | $\Phi_{18} = a_1 J_1 + a_{11} J_1^2 + a_{111} J_1^3 + a_{1111} J_1^4 + a_{11111} J_1^5 + a_2 J_2 + a_3 J_3 + a_{12} J_1 J_2 + a_{112} J_1^2 J_2$ | 10 |
| $a_1$, $a_2$, $a_{11}$, $a_{111}$, $a_{12}$, $a_3$ | $\Phi_{19} = \Phi_5 + a_{1111} J_1^4 + a_{13} J_1 J_3 + a_{113} J_1^2 J_3 + a_{122} J_1 J_2^2 + a_{23} J_2 J_3 + a_{1122} J_1^2 J_2^2 + a_{123} J_1 J_2 J_3$ | 12 |
|  | $\Phi_{20} = \Phi_5 + a_{1111} J_1^4 + a_{13} J_1 J_3 + a_{113} J_1^2 J_3 + a_{122} J_1 J_2^2 + a_{11111} J_1^5 + a_{1122} J_1^2 J_2^2 + a_{1113} J_1^3 J_3$ | 12 |
|  | $\Phi_{21} = \Phi_5 + a_{1111} J_1^4 + a_{13} J_1 J_3 + a_{113} J_1^2 J_3 + a_{23} J_2 J_3 + a_{11111} J_1^5 + a_{11112} J_1^4 J_2 + a_{123} J_1 J_2 J_3$ | 12 |
|  | $\Phi_{22} = \Phi_5 + a_{1111} J_1^4 + a_{13} J_1 J_3 + a_{113} J_1^2 J_3 + a_{122} J_1 J_2^2 + a_{1112} J_1^3 J_2 + a_{1122} J_1^2 J_2^2 + a_{1113} J_1^3 J_3$ | 12 |
|  | $\Phi_{23} = \Phi_5 + a_{1111} J_1^4 + a_{13} J_1 J_3 + a_{23} J_2 J_3 + a_{122} J_1 J_2^2 + a_{11111} J_1^5 + a_{222} J_2^3 + a_{33} J_3^2$ | 12 |

The Landau-Devonshire potentials shown in the Table 1 were obtained by truncating of the unessential terms without change of the singularities of the potential (4), and they can be widely used in the phenomenological studies of the various crystals including perovskite crystals.

### III. HIGHER ORDER LANDAU-DEVONSHIRE POTENTIAL MODEL FOR BaTiO3

Having discussed the phenomenological theory on the temperature dependence of various coefficients of the previous potentials (7) and (11), and the first principles calculation[28], we assume three temperature dependent coefficients $a_1$, $a_{11}$ and $a_2$ in the potential (4).

Hence, we have to find the universal unfolding of the singularity of the potential with three control parameters to obtain the structurally stable potential model with variable parameters $a_1$, $a_{11}$ and $a_2$.

According to the Table I, such structurally stable potential model has to include at least all the invariants up to the eighth power:

$$\Phi = a_1(T)J_1 + a_{11}(T)J_1^2 + a_2(T)J_2 + a_{111}J_1^3 + a_3J_3 + a_{12}J_1J_2 + a_{1111}J_1^4 + a_{112}J_1^2J_2 + a_{22}J_2^2 + a_{13}J_1J_3 \quad (15)$$

We will add two tenth order invariants $J_1^2J_3$ and $J_2J_3$ to the model (15) for qualitative and quantitative description of the recent experimental data[27, 33] on BaTiO$_3$. We then obtain the following potential:

$$\tilde{\Phi} = a_1(T)J_1 + a_{11}(T)J_1^2 + a_2(T)J_2 + a_{111}J_1^3 + a_3J_3 + a_{12}J_1J_2 + \\ + a_{1111}J_1^4 + a_{112}J_1^2J_2 + a_{22}J_2^2 + a_{13}J_1J_3 + a_{113}J_1^2J_3 + a_{23}J_2J_3 \quad (16)$$

Here $a_1$, $a_{11}$ and $a_2$ are assumed to be temperature dependent linearly:

$$\begin{cases} a_1 = a_1^0(T-T_0) = a_1^0(T-113) \\ a_{11} = a_{11}^0(T-T_0) + b = a_{11}^0 T + a_{11}^1 , \\ a_2 = a_2^0(T-T_0) + c = a_2^0 T + a_2^1 \end{cases} \quad (17)$$

and $a_1$ obeys the Curie-Weiss law

$$a_1 = \frac{T-T_0}{2\varepsilon_0 C_0},$$

where $C_0$ is the Curie-Weiss constant; $T_0$ is the Curie-Weiss temperature; $\varepsilon_0$ is the dielectric constant of vacuum and $T$ is temperature,

The experimental data on the cubic-tetragonal phase transition temperature, spontaneous polarization[33], dielectric constant[27] and piezoelectric coefficients[27] in the tetragonal phase *P4mm* are used to obtain the values of the coefficients $a_1$, $a_{11}$, $a_{111}$ and $a_{1111}$. And the values of the coefficients $a_2$, $a_{12}$, $a_{112}$ and $a_{22}$ are determined from the experimental data associated with the ferroelectric orthorhombic phase *Amm*2[33], and the coefficients $a_3$, $a_{13}$, $a_{113}$ and $a_{23}$ -the ferroelectric properties in the rhombohedral phase *R*3*m*[33].

In Table II, we list values of the expansion coefficients of the higher Landau-Devonshire potential model (16).

TABLE II. The Coefficients of the Landau-Devonshire potential (16) for BaTiO$_3$

| Coeffi-cients | This work | Notations from Refs. | | | | Units |
|---|---|---|---|---|---|---|
| | | 6 | 20 | 7 | 8 | |
| $a_1$ | $5.6497 \times 10^5 \times$ ($T$-113) | $3.34 \times 10^5 \times$ ($T$-108) | $3.33 \times 10^5 \times$ ($T$-110) | $4.124 \times 10^5 \times$ ($T$-115) | $3.16 \times 10^5 \times$ ($T$-118) | Vm C$^{-1}$ |
| $a_{11}$ | $-1.1377 \times 10^6 T+$ $+1.9162 \times 10^8$ | $-2.045 \times 10^9+$ $+4.69 \times 10^6 T$ | $3.6 \times 10^6 \times$ ($T$-175) | $-2.097 \times 10^8$ | $-1.83 \times 10^9+$ $+4 \times 10^6 T$ | Vm$^5$C$^{-3}$ |
| $a_2$ | $9.5072 \times 10^5 T-$ $-2.3230 \times 10^8$ | $3.23 \times 10^8$ | $4.9 \times 10^8$ | $7.974 \times 10^8$ | $-2.24 \times 10^9+$ $+6.7 \times 10^6 T$ | Vm$^5$C$^{-3}$ |
| $a_{111}$ | $4.3206 \times 10^9$ | $2.445 \times 10^{10}-$ $-5.52 \times 10^7 T$ | $6.6 \times 10^9$ | $1.294 \times 10^9$ | $1.39 \times 10^{10}-$ $-3.2 \times 10^7 T$ | Vm$^9$C$^{-5}$ |
| $a_{12}$ | $1.4584 \times 10^9$ | $4.47 \times 10^9$ | $2.9 \times 10^9$ | $-1.950 \times 10^9$ | $-2.2 \times 10^9$ | Vm$^9$C$^{-5}$ |
| $a_3$ | $9.1543 \times 10^{10}$ | $4.91 \times 10^9$ | $7.6 \times 10^7 \times$($T$-120)+$4.4 \times 10^9$ | $-2.500 \times 10^9$ | $5.51 \times 10^{10}$ | Vm$^9$C$^{-5}$ |
| $a_{1111}$ | $1.6378 \times 10^{10}$ | | | $3.863 \times 10^{10}$ | $4.84 \times 10^{10}$ | Vm$^{13}$C$^{-7}$ |
| $a_{112}$ | $-3.8696 \times 10^{10}$ | | | $2.529 \times 10^{10}$ | $2.53 \times 10^{11}$ | Vm$^{13}$C$^{-7}$ |
| $a_{22}$ | $-1.0070 \times 10^{11}$ | | | $1.637 \times 10^{10}$ | $2.80 \times 10^{11}$ | Vm$^{13}$C$^{-7}$ |
| $a_{13}$ | $-1.0346 \times 10^{12}$ | | | $1.367 \times 10^{10}$ | $9.35 \times 10^{10}$ | Vm$^{13}$C$^{-7}$ |
| $a_{113}$ | $2.4318 \times 10^{12}$ | | | | | Vm$^{17}$C$^{-9}$ |
| $a_{23}$ | $8.1241 \times 10^{11}$ | | | | | Vm$^{17}$C$^{-9}$ |

# IV. TEMPERATURE AND ELECTRIC FIELD DEPENDENCE OF FERROELECTRIC PROPERTIES IN BaTiO$_3$

## A. Temperature dependence of spontaneous polarization

The spontaneous polarization of the low temperature phases is determined by the solutions of the state equation $\partial \Phi / \partial p_i = 0$. The tetragonal structure has $p_1 = p_2 = 0$, $p_3 \neq 0$, the orthorhombic one - $p_2 = 0$, $p_1 = p_3 \neq 0$, and the rhombohedral one - $p_1 = p_2 = p_3 \neq 0$. Then, using the tenth potential model (16), we obtain following state equations for three ferroelectric phases:

For the tetragonal phase,
$$4a_{1111} p_3^6 + 3a_{111} p_3^4 + 2a_{11} p_3^2 + a_1 = 0.$$

For the orthorhombic phase,
$$2(16 a_{1111} + 4 a_{112} + a_{22}) p_3^6 + 3(4 a_{111} + a_{12}) p_3^4 + (4 a_{11} + a_2) p_3^2 + a_1 = 0.$$

For the rhombohedral phase,
$$5(3 a_{113} + a_{23}) p_3^8 + 4(27 a_{1111} + 9 a_{112} + 3 a_{22} + a_{13}) p_3^6 +$$
$$+ (27 a_{111} + 9 a_{12} + a_3) p_3^4 + 2(3 a_{11} + a_2) p_3^2 + a_1 = 0.$$

The spontaneous polarizations versus temperature determined from above state equations are shown in Fig. 1(a).

## B. Temperature dependence of dielectric constants

The inverse dielectric susceptibility tensor is determined by the Hessian of the potential model (16):
$$\left\| \chi_{ij}^{-1} \right\| = \left\| \partial^2 \Phi / \partial p_i \partial p_j \right\|$$

For the tetragonal phase, the elements of the dielectric susceptibility tensor are:
$$\chi_{11T}^{-1} = \chi_{22T}^{-1} = 2 a_{112} p_3^6 + 2 a_{12} p_3^4 + 2 a_2 p_3^2$$
$$\chi_{33T}^{-1} = 48 a_{1111} p_3^6 + 24 a_{111} p_3^4 + 8 a_{11} p_3^2$$
$$\chi_{12T}^{-1} = \chi_{21T}^{-1} = \chi_{13T}^{-1} = \chi_{31T}^{-1} = \chi_{23T}^{-1} = \chi_{32T}^{-1} = 0.$$

For the orthorhombic phase,
$$\chi_{11O}^{-1} = \chi_{33O}^{-1} = 8(24 a_{1111} + 5 a_{112} + a_{22}) p_3^6 + 8(6 a_{111} + a_{12}) p_3^4 + 8 a_{11} p_3^2$$
$$\chi_{22O}^{-1} = 2(4 a_{113} + a_{23}) p_3^8 + 4(2 a_{112} + a_{13} + a_{22}) p_3^6 + 2(2 a_{12} + a_3) p_3^4 + 2 a_2 p_3^2$$
$$\chi_{13O}^{-1} = \chi_{31O}^{-1} = 8(24 a_{1111} + 7 a_{112} + 2 a_{22}) p_3^6 + 16(3 a_{111} + a_{12}) p_3^4 + 4(2 a_{11} + a_2) p_3^2$$
$$\chi_{12O}^{-1} = \chi_{21O}^{-1} = \chi_{23O}^{-1} = \chi_{32O}^{-1} = 0.$$

For the rhombohedral phase,
$$\chi_{11R}^{-1} = \chi_{22R}^{-1} = \chi_{33R}^{-1} = 8(7 a_{113} + 2 a_{23}) p_3^8 + 8(54 a_{1111} + 15 a_{112} + 4 a_{22} + a_{13}) p_3^6 +$$
$$+ 8(9 a_{111} + 2 a_{12}) p_3^4 + 8 a_{11} p_3^2$$
$$\chi_{12R}^{-1} = \chi_{21R}^{-1} = \chi_{13R}^{-1} = \chi_{31R}^{-1} = \chi_{23R}^{-1} = \chi_{32R}^{-1} = 4(23 a_{113} + 8 a_{23}) p_3^8 +$$
$$+ 4(108 a_{1111} + 39 a_{112} + 14 a_{22} + 5 a_{13}) p_3^6 + 4(18 a_{111} + 7 a_{12} + a_3) p_3^4 + 4(2 a_{11} + a_2) p_3^2.$$

We can set a new coordinate system [$x_1'$, $x_2'$, $x_3'$] to let the new $[001]'$ direction parallel to the polarization direction in each ferroelectric phase. The relations of dielectric susceptibility between the old and new coordinates are given below.

For the tetragonal phase:
$$\chi_{11T}'^{-1} = \chi_{22T}'^{-1} = \chi_{11T}^{-1}, \quad \chi_{33T}'^{-1} = \chi_{33T}^{-1},$$

$$\chi'^{-1}_{ijT} = 0 \, (i \neq j).$$

For the orthorhombic phase:
$$\chi'^{-1}_{11O} = \chi^{-1}_{11O} - \chi^{-1}_{13O}, \quad \chi'^{-1}_{22O} = \chi^{-1}_{22O}, \quad \chi'^{-1}_{33O} = \chi^{-1}_{11O} + \chi^{-1}_{13O},$$
$$\chi'^{-1}_{ijO} = 0 \, (i \neq j).$$

For the rhombohedral phase:
$$\chi'^{-1}_{11R} = \chi'^{-1}_{22R} = \chi^{-1}_{11R} - \chi^{-1}_{12R}, \quad \chi'^{-1}_{33R} = \chi^{-1}_{11R} + 2\chi^{-1}_{12R},$$
$$\chi'^{-1}_{ijR} = 0 \, (i \neq j).$$

The dielectric constants $\varepsilon_{ij}$ are calculated from the dielectric susceptibility by
$$\varepsilon'_{ij} = \frac{1}{\varepsilon_0} \chi'^{-1}_{ij}.$$

The calculated results are shown in Fig. 1(b).

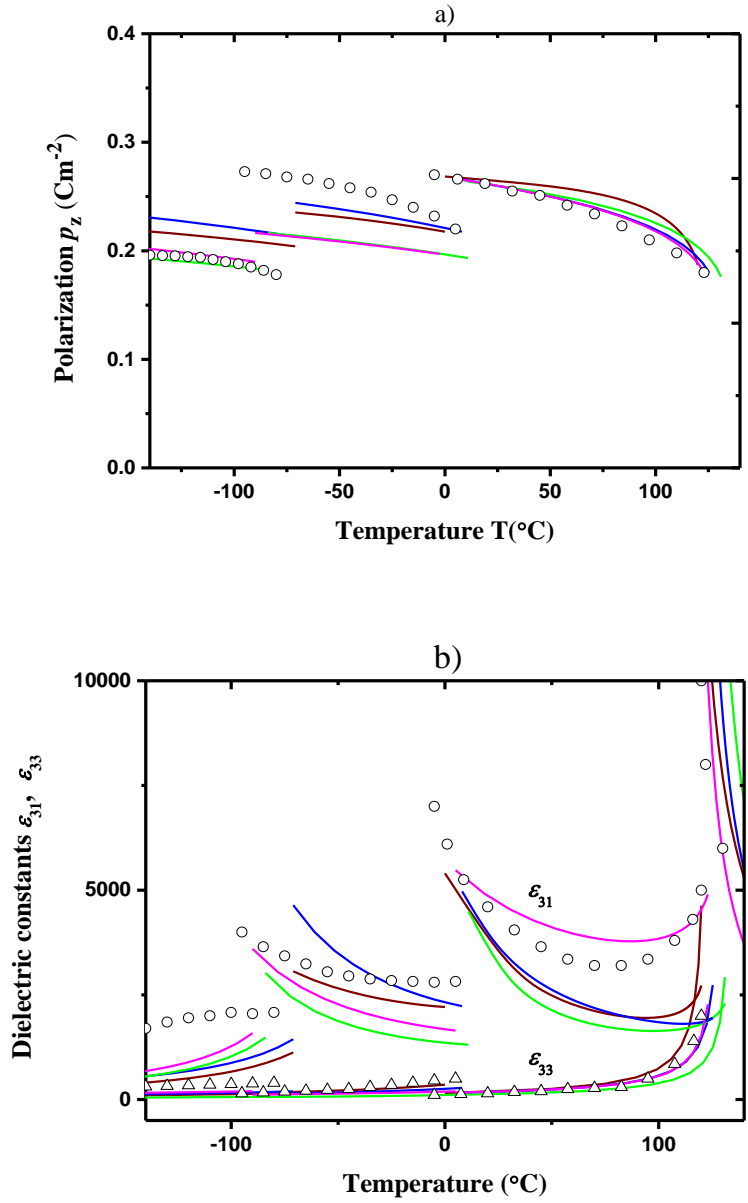

Fig. 1. Calculation results on (a) the spontaneous polarization versus temperature and (b) the dielectric constants versus temperature in BaTiO$_3$(pink lines). The previous experimental and phenomenological results are presented for comparison with the calculation results. Experimental data, denoted by ○'s and △'s[W. Martienssen and H.Warlimont, editors. *Springer Handbook of Condensed Matter and Materials Data*, (Springer Berlin Heidelberg 2005)]. Theoretical data, denoted by solid lines: brown lines are from A. J. Bell, J. Appl. Phys. **89**, 3907(2001), blue from Y. L. Li, L. E. Cross, and L. Q. Chen, J. Appl. Phys. **98**, 064101 (2005), and green from Y. L. Wang, A. K. Tagantsev, D. Damjanovic, and N. Setter, J. Appl. Phys. **101**, 104115 (2007), respectively.

### C. Ferroelectric properties of the tetragonal phase under the applied electric field $E_{[001]} = E_0$

The potential for description of the physical properties of BaTiO$_3$ crystal under the applied electric field is

$$\Phi = \tilde{\Phi}(J) - E_0 p_3. \tag{18}$$

Using the potential (18), we can obtain the spontaneous polarization($p_3$), dielectric susceptibility($\chi_{33}$), piezoelectric coefficients($d_{31}, d_{33}$) for tetragonal phase under the applied electric field $E_{[001]}$ at the room temperature(T=25℃).

The spontaneous polarization is determined from

$$8a_{1111}p_3^7 + 6a_{111}p_3^5 + 4a_{11}p_3^3 + 2a_1 p_3 - E_0 = 0. \tag{19}$$

The dielectric susceptibility is

$$\chi_{33} = (56a_{1111}p_3^6 + 30a_{111}p_3^4 + 12a_{11}p_3^2 + 2a_1)^{-1} \tag{20}$$

The piezoelectric coefficients are

$$d_{31} = \frac{de_{[100]}}{dE_{[100]}} = 2Q_{12}p_3\chi_{33} (= d_{32}), \quad d_{33} = \frac{de_{[001]}}{dE_{[001]}} = 2Q_{11}p_3\chi_{33}, \tag{21}$$

where $e_{[lmn]}$ is the strain under zero stress. The strains are calculated by

$$e_{[lmn]} = e_{11}^0 l^2 + e_{22}^0 m^2 + e_{33}^0 n^2 + 2e_{12}^0 lm + 2e_{13}^0 ln + 2e_{23}^0 mn$$

$$e_{11}^0 = Q_{11}p_1^2 + Q_{12}p_2^2 + Q_{12}p_3^2, \quad e_{12}^0 = Q_{44}p_1 p_2$$

$$e_{22}^0 = Q_{12}p_1^2 + Q_{11}p_2^2 + Q_{12}p_3^2, \quad e_{13}^0 = Q_{44}p_1 p_3$$

$$e_{33}^0 = Q_{11}p_1^2 + Q_{12}p_2^2 + Q_{13}p_3^2, \quad e_{23}^0 = Q_{44}p_2 p_3$$

where $l, m, n$ is the direction unit vector with $l^2 + m^2 + n^2 = 1$.
In Fig. 2(a)-(c), the results calculated by (19)-(21) are shown, respectively.

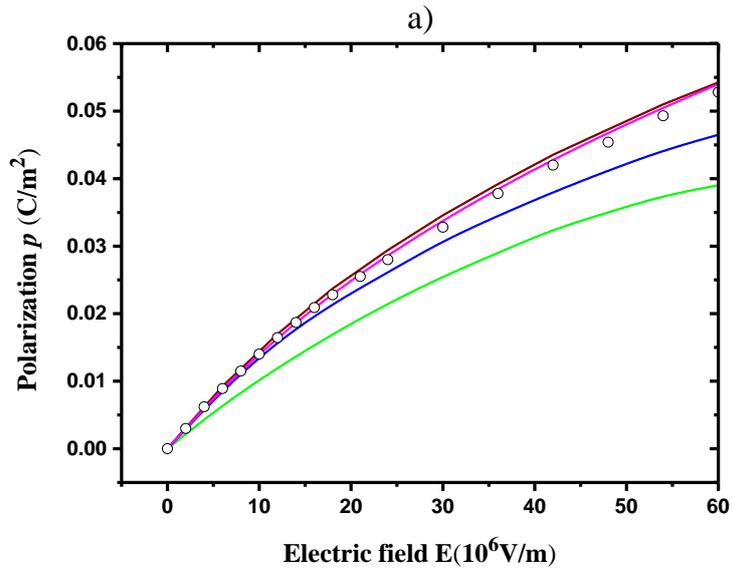

a)

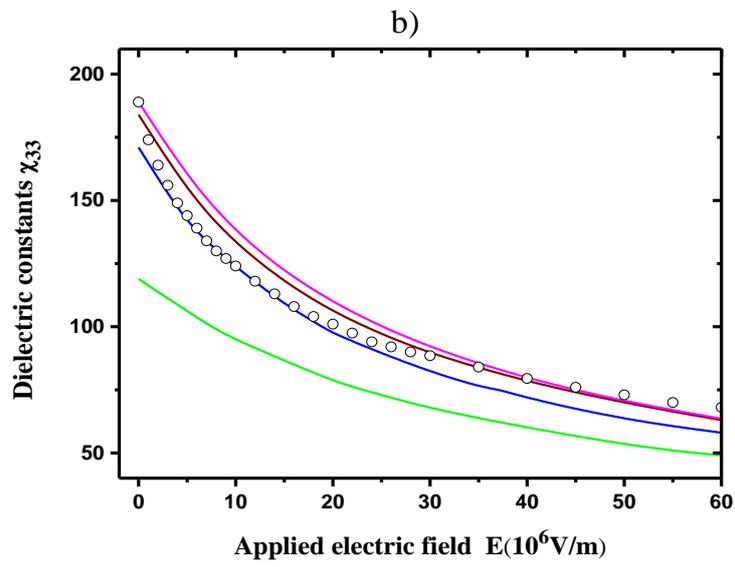

b)

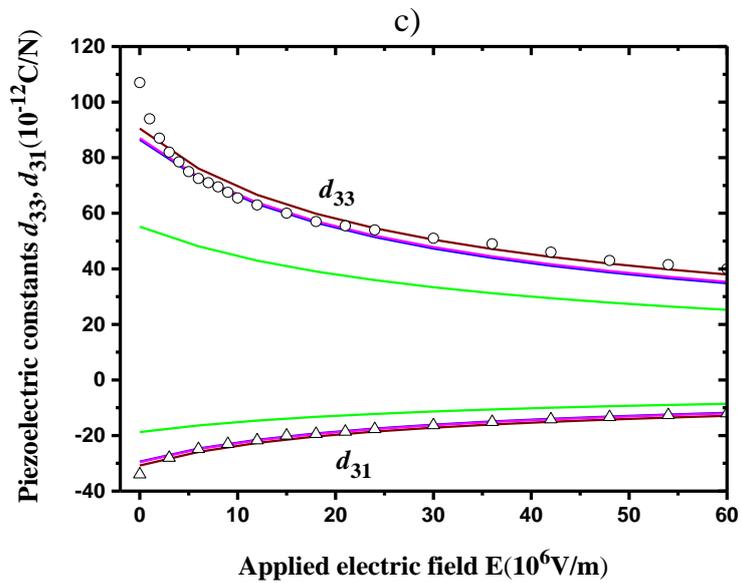

c)

Fig. 2. Calculation results on the electric field dependences of (a) the spontaneous polarization $p_3$, (b) the dielectric susceptibility $\chi_{33}$ and (c) the piezoelectric coefficients $d_{31}$, $d_{33}$ at room temperature(T=25°C) in BaTiO$_3$(pink lines). The previous experimental and phenomenological results are presented for comparison with the calculation results. Experimental data, denoted by ○'s and △'s[I. N. Leontyev, O. E. Fesenko, N. G. Leontyev, and B. Dkhil, Appl. Phys. Lett. **96**, 142904 (2010)]. Theoretical data, denoted by solid lines: brown lines are from A. J. Bell, J. Appl. Phys. **89**, 3907(2001), blue from Y. L. Li, L. E. Cross, and L. Q. Chen, J. Appl. Phys. **98**, 064101 (2005), and green from Y. L. Wang, A. K. Tagantsev, D. Damjanovic, and N. Setter, J. Appl. Phys. **101**, 104115 (2007), respectively.

## IV. DISCUSSION

The sixth order potentials by Bell et al.[5] and Pertsev et al.[20] and the eighth order ones by Y. L. Wang et al.[8] and J. J. Wang et al.[12] are structurally unstable from the singularity theory. These structurally unstable models in the previous phenomenological theories gave the agreement with the experiments in some range of the external parameters. They have great popularity due to simplicity. However, these models always yield incorrect results. Any small change in the external conditions, i.e. a perturbation, results in a qualitative change in the results. Therefore, it is obvious that the qualitative characteristics of the phase transitions in BaTiO$_3$ crystal cannot be precisely described by the phenomenological theories based on these potentials.

Although the eighth order potential (11) by Li et al.[5] is structurally stable and the kind and number of the invariants contained in the potential are the same as those in the potential (15), the singularities of these two potentials are different. The potential (11) corresponds to the universal unfolding of the germ with one control parameter, while the potential (15) - three control parameters. Therefore, the potentials (11) and (15) are different potential models.

In the case of the potential (15), we can consider the phase transitions occurring on the thermodynamic paths along all possible directions in the three dimensional phenomenological coefficient parameter space ($a_1 - a_{11} - a_2$), while in the case of the potential (11) - along the only one direction in one dimensional $a_1$ space. That is, we can only describe the phase transition along the temperature axis parallel to the $a_1$ axis

But in the potential (15), coefficients $a_1$, $a_{11}$ and $a_2$ depend on the temperature, and it is possible to describe better the characteristics of the phase transitions. We can investigate the phase transitions with change of three external parameters such as temperature and pressure etc. by potential (15) unlike potential (11). After all, the description of the experimental data on the phase transition in BaTiO$_3$ crystal by potential (15) includes more general and universal results than previous potential models provide.

The tenth order potential model (16) has been proposed by adding of the tenth order invariants $J_1^2 J_3$ and $J_2 J_3$ to the potential model (15) to describe better the experimental data observed in the rhombohedral phase. These two tenth order invariants don't affect the structural stability of the potential (15), because they belong to the gradient ideal $I_{\nabla \Phi}$.

The changes of the ferroelectric properties of BaTiO$_3$ crystal under zero electric field over the whole range of temperatures in which spontaneous polarization appear are shown in Fig. 1. The Fig. 2 shows the dielectric and piezoelectric properties of the crystal under superstrong electric field $E_{[001]}$ ~55MV/m.

We have compared the results of the phenomenological theory based on the new tenth order Landau-Devonshire potential (16) with previous theoretical and experimental results (Fig. 1, 2).

The experimental data of the spontaneous polarization versus electric field are fitted better by Bell et al.'s and our models rather than Wang et al.'s and Li et al.'s ones.

The experiments on the electric field dependence of the dielectric susceptibility are in good agreement with the results of Bell et al.'s and our models. On the other hand, the dependences of the piezoelectric coefficients on the electric field are explained well quantitatively as well as qualitatively by Li et al.'s and our models.

From the figures, it is known that our results based on the potential (16) are in better agreement with the recent experimental data[27, 33] than those reported in the Ref. [6-8]. It is found that our results explain qualitatively and quantitatively the experimental data on the dependences of the spontaneous polarization, dielectric constant, piezoelectric coefficients of the ferroelectric phase $P4mm$ under superstrong electric field $E_{[001]}$~55MV/m. (Fig. 2)

## V. SUMMARY

We have shown that the Landau-Devonshire potentials by A. J. Bell et al.[5], N. A. Pertsev et al.[20], Y. L. Wang et al.[8] and J. J. Wang et al.[12] widely used in phenomenology on $BaTiO_3$ crystal are structurally unstable.

We have obtained 23 structurally stable Landau potential with Co-dim=1, 2, 3, 4, 6 able to be used in the phenomenological study of the phase transitions in the perovskite-type crystals. (Table 1)

The structurally stable Landau-Devonshire potential with three variable parameters different from the Li et al.'s model[5] with one variable parameter is proposed for phenomenological description of $BaTiO_3$ crystal.

With the new tenth potential model (16), we calculated the temperature and electric field dependences of the spontaneous polarization, dielectric constants, piezoelectric coefficients of $BaTiO_3$ crystal. We believe that our data is in good agreement with experimental data, despite the slight differences. The results show that the tenth potential (16) describes more precisely the qualitative and quantitative characteristic of the recent experimental data than those of the previous phenomenological studies.

---------------------------------------------------------------------------

1. D. Vanderbilt and M.H. Cohen, Phys. Rev., **B 63**, 094108 (2001).
2. I. A. Sergienko, Yu. M. Gufan, and S. Urazhdin, Phys. Rev. B **65**, 144104 (2002).
3. L. Li, I. H. Kim, K. O. Jang, K. S. Ri, and J. C. Cha, J. Appl. Phys. **114**, 034104 (2013).
4. I. H. Kim, K.O. Jang, I. H. Kim, K. S. Ri, and C.S. Xu, Physica B **424**, 20 (2013).
5. A. J. Bell and L. E. Cross, Ferroelectrics **59**, 197 (1984).
6. A. J. Bell, J. Appl. Phys. **89**, 3907(2001).
7. Y. L. Li, L. E. Cross, and L. Q. Chen, J. Appl. Phys. **98**, 064101 (2005).
8. Y. L. Wang, A. K. Tagantsev, D. Damjanovic, N. Setter, V. K. Yarmarkin, A. I. Sokolov and I. A. Lukyanchuk, J. Appl. Phys. **101**, 104115 (2007).
9. L. Liang, Y. L. Li, L. Q. Chen, S. Y. Hu and G. H. Lu, J. Appl. Phys. **106**, 104118, (2009).
10. Y. L. Wang, A. K. Tagantsev, D. Damjanovic, N. Setter, V. K. Yarmarkin, and A. I. Sokolov, Phys. Rev. B **73**, 132103 (2006).
11. L. Liang, Y. L. Li, L.-Q. Chen, S. Y. Hu, and G.-H. Lu, Appl. Phys. Lett. **94**, 072904 (2009).
12. J. J. Wang, P. P. Wu, X. Q. Ma, and L. Q. Chen, J. Appl. Phys. **108**, 114105, (2010).
13. X. Lu, H. Li, and W. Cao, J. Appl. Phys. **114**, 224106 (2013).
14. G. Gaeta, Ann. Phys., 312, 511 (2004).
15. P. Toledano and V. Dmitriev, *Reconstructive phase transitions in crystals and quasicrystals* (World Scientific, Singapore, 1996).
16. E. I. Kutin, V. L. Lorman, and S. V. Pavlov, Sov. Phys. Usp. **34**, 497 (1991).

17. V. I. Arnol'd, A. N. Varchenko, and S. M. Gusein-Zade, *Singularities of Differentiable Maps*, Vol. 1, 2 (Birkhauser, Boston, 1985, 1988).
18. V. I. Arnold, editor. *Dynamical Systems* V: *Bifurcation Theory and Catastrophe Theory*, volume 5 of *Encyclopaedia of Mathematical Sciences* (Springer-Verlag, Berlin, 1994).
19. T. Poston and I. N. Stewart, *Catastrophe theory and its applications* (New York, Dover, 1996).
20. N. A. Pertsev, A. G. Zembilgotov, and A. K. Tagantsev, Phys. Rev. Lett. **80**, 1988 (1998).
21. V. J. Koukhar, N. A. Pertsev, and R. Waser, Phys. Rev. B **64**, 214103 (2001).
22. D. A. Tenne, X. X. Xi, Y. L. Li, L. Q. Chen, A. Soukiassian, M. H. Zhu, A. R. James, J. Lettieri, D. G. Schlom, W. Tian, and X. Q. Pan, Phys. Rev. B **69**, 174101 (2004).
23. K. J. Choi, M. Biegalski, Y. L. Li, A. Sharan, J. Schubert, R. Uecker, P. Reiche, Y. B. Chen, X. Q. Pan, V. Gopalan, L.-Q. Chen, D. G. Schlom, and C. B. Eom, Science **306**, 1005 (2004).
24. V. B. Shirokov, Yu. I. Yuzyuk, B. Dkhil, and V. V. Lemanov, Phys. Rev. B **79**, 144118, 2009.
25. V. B. Shirokov, Yu. I. Yuzyuk, B. Dkhil, and V. V. Lemanov, Phys. Rev. B **75**, 224116 (2007).
26. V. B. Shirokov, V. I. Torgashev, A. A. Bakirov, and V. V. Lemanov, Phys. Rev. B **73**, 104116 (2006).
27. I. N. Leontyev, O. E. Fesenko, N. G. Leontyev, and B. Dkhil, Appl. Phys. Lett. **96**, 142904 (2010).
28. A. I. Sokolov, Fiz. Tverd. Tela 51, 2, (2009). (in Russian) [Phys. Solid Stat. 51, 2, 351 (2009)].
29. Yu. M. Gufan, *Structural Phase Transitions* (in Russian), (Nauka, M., 1982).
30. Yu. A. Izyumov and V. N. Syromiatnikov, *Phase transitions and crystal symmetry* (Dordrecht: Kluwer Academic Publishers 1990).
31. M. E. Lines and A. M. Glass, *Principles and Applications of Ferroelectrics and Related Materials* (Oxford University Press, New York, 1977).
32. K. M. Rabe, C. H. Ahn, and J.-M. Triscone, editors. *Physics of Ferroelectrics*: *A Modern Perspective* (Springer-Verlag, Berlin, Heidelberg 2007).
33. W. Martienssen and H.Warlimont, editors. *Springer Handbook of Condensed Matter and Materials Data*, (Springer Berlin Heidelberg 2005).